\documentclass{LT23auth}
\usepackage{graphicx}
\usepackage{amssymb}
\newcommand{\vell}{\mbox{\boldmath$\ell$}}
\def\vJ{{\bf J}}
\def\vB{{\bf B}}
\begin{document}
\begin{frontmatter}
\title{Thermal conductivity of superfluid $^3$He in aerogel}
\author{Priya Sharma}
\author{and J.A.Sauls\thanksref{thank1}} 
\address{Department of Physics and Astronomy, Northwestern University,Evanston IL 60208, USA}
\thanks[thank1]{ sauls@snowmass.phys.nwu.edu}
\begin{abstract}
We report theoretical calculations of the thermal conductivity of superfluid $^3$He impregnated 
into high-porosity aerogel and compare these results with available experimental data.
\end{abstract}
\end{frontmatter}
When $^3$He is impregnated into high-porosity aerogel, a new scattering channel is available
to $^3$He quasiparticles, viz., elastic scattering off the aerogel strands. We examine the
effects of elastic and inelastic scattering on the transport properties of $^3$He in aerogel,
and report new results for the thermal conductivity of superfluid $^3$He in aerogel within the
framework of homogeneous and isotropic scattering. This model predicts significant
variations in the temperature dependence of the thermal conductivity as a function of
pressure, scattering cross-section and aerogel density. Measurements of the thermal conductivity
of $^3$He in $98\,\%$ aerogel at $p=7.4\,\mbox{bar}$ \cite{fis02} are in good agreement with theoretical
calculations based on either the BW or the ABM phase order parameters. At higher pressures,
where pairbreaking effects are weaker, significant differences in the thermal conductivity for
these two phases are predicted.

\begin{figure}[!tbp]
\begin{center}\leavevmode
\includegraphics[width=0.8\linewidth]{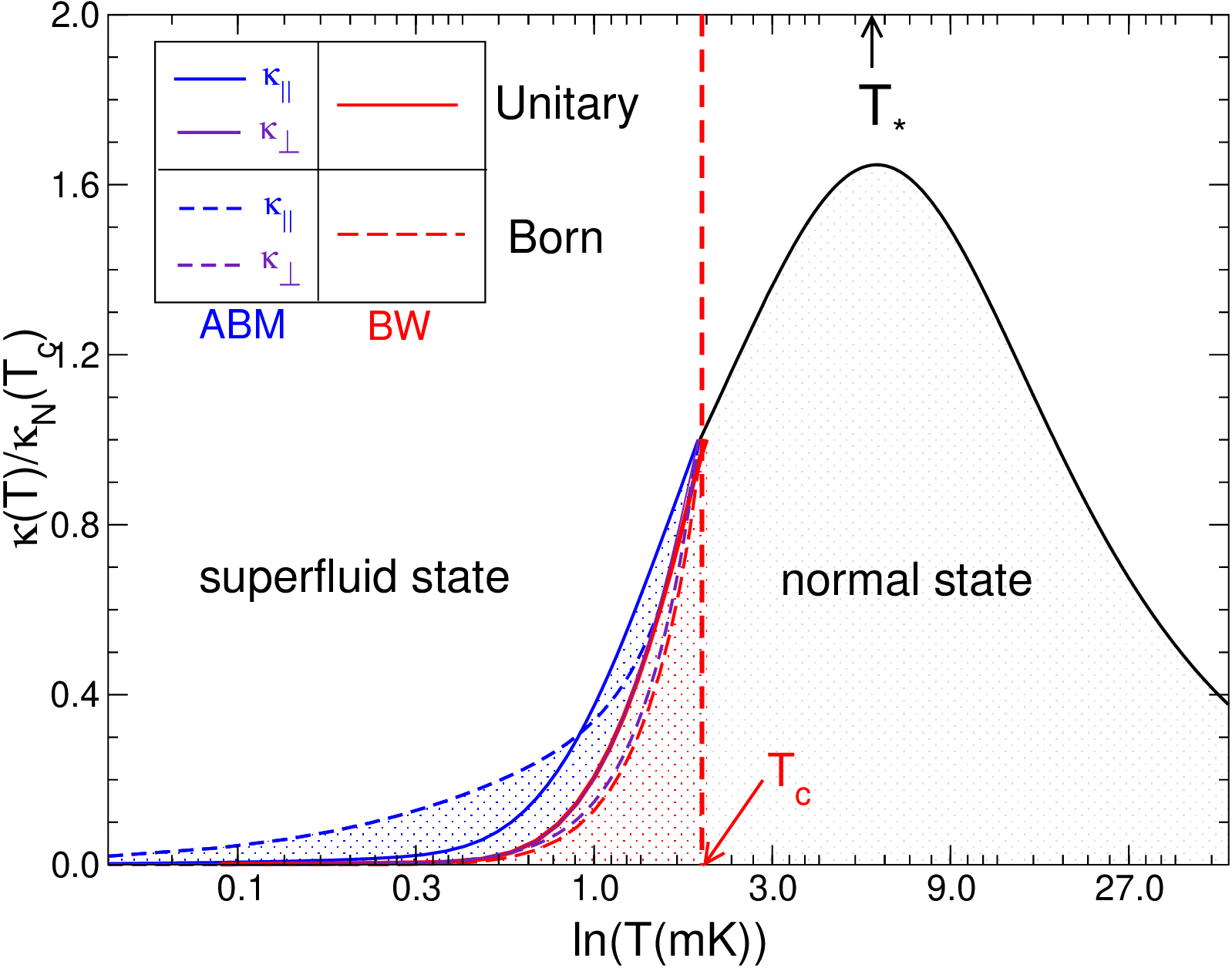}
\caption{Theoretical calculations of $\kappa$ at $p=30$ bar for a mean free path of $\ell=180\,\mbox{nm}$ in 
         both the Born and Unitary limits.}
\label{Kappa_vs_LogT}
\end{center}
\end{figure}

Figure \ref{Kappa_vs_LogT} summarizes theoretical calculations for the thermal conductivity at
a pressure of $p=30\,\mbox{bar}$ over a temperature range, $0.01\,\mbox{mK}<T\le 30\,\mbox{mK}$.
At high temperatures $T>T_{*}$ in the normal-state the transport mean-free path is determined by 
quasiparticle-quasiparticle scattering. Thus, we recover the bulk thermal conductivity of pure
$^3$He with $\kappa\propto 1/T$ for $T>T_{*}$ ($\approx 5.5\,\mbox{mK}$ at this pressure).
The thermal conductivity crosses over in the normal state to a low-temperature regime 
determined by elastic scattering as shown in Fig. \ref{Kappa_vs_LogT}. 
This calculation is based on an exact solution to the Boltzmann-Landau transport equation with both elastic 
and inelastic scattering \cite{sha00}. For $T\ll T_*$ in the normal-state $\kappa\propto T$ and 
the mean-free path in the aerogel, $\kappa=\frac{\pi^2}{9}N_f k_B^2 T v_f\ell$. 
For aerogel with a porosity of $98\%$ we estimate $\ell\approx 180\,\mbox{nm}$ \cite{thu98}.

\begin{figure}[tbp!]
\begin{center}\leavevmode
\includegraphics[width=0.95\linewidth]{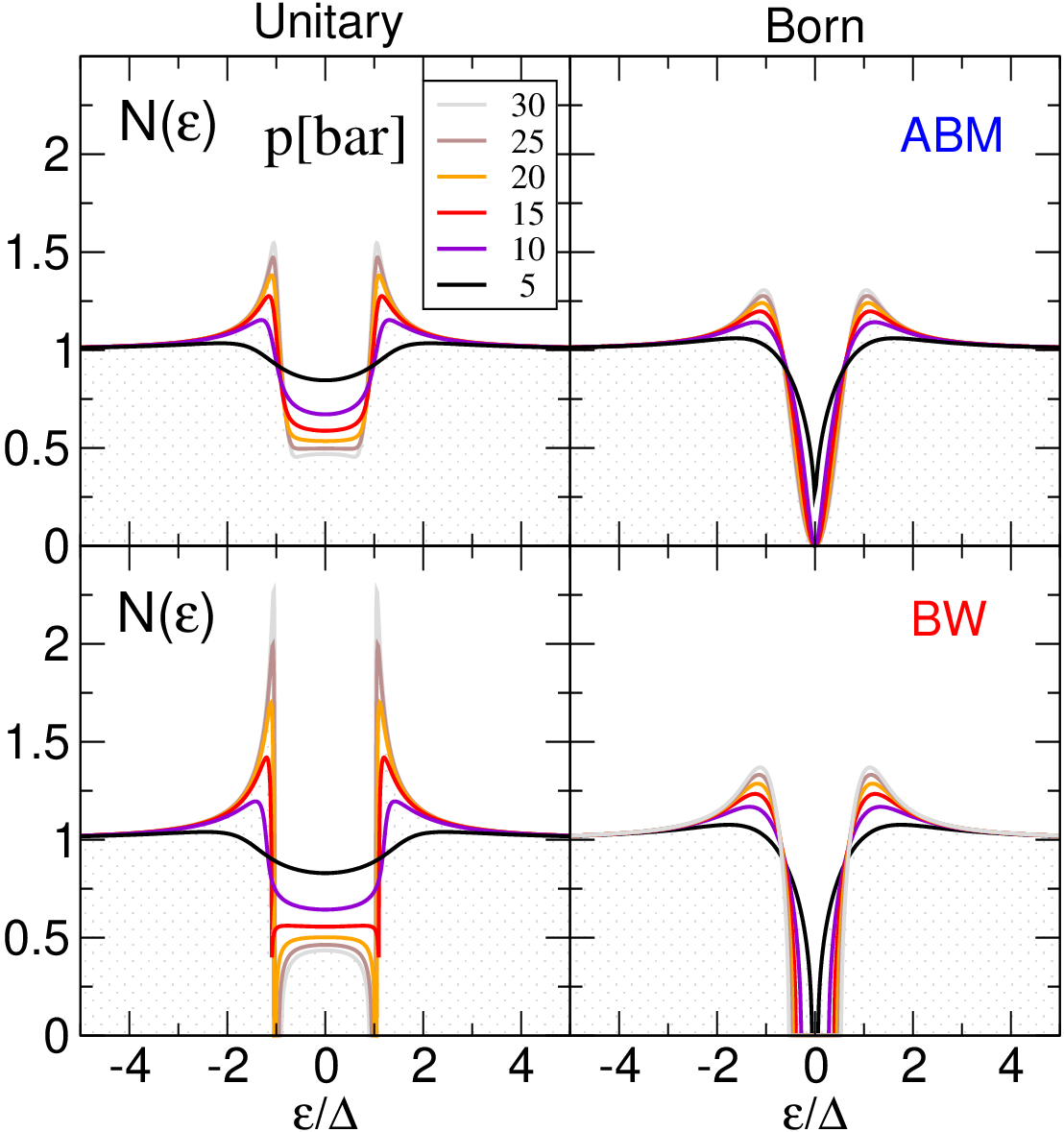}
\caption{Theoretical calculations of the density of states for ABM- and BW
order parameters with aerogel scattering. The mean-free path is $\ell=180\,\mbox{nm}$
and the scattering is in both the unitary and Born limits. The spectra are calculated
at reduced temperatures, $T/T_{ca}=0.2$.}
\label{DOS}
\end{center}
\end{figure}

Scattering by the aerogel matrix leads to a suppression of the superfluid transition, pairbreaking
\cite{thu98} and the formation of a spectrum of low-energy quasiparticle states below the continuum 
gap edge. The spectrum of these excitations generally depends upon the symmetry of the order parameter, as 
well as the scattering cross-section and mean-free path \cite{gra96a} (see Fig. \ref{DOS}). In the unitary 
limit (strong scattering) a band of gapless excitations {\sl at the Fermi level}, with energies 
$|\varepsilon|\le\gamma\approx0.67\Delta\sqrt{\xi_0/\ell}$, forms which is relatively insensitive to 
the symmetry of the order parameter, particularly at lower pressures where $\xi_0/\ell$ is largest. 

\begin{figure}[ht!]
\begin{center}\leavevmode
\includegraphics[width=0.850\linewidth]{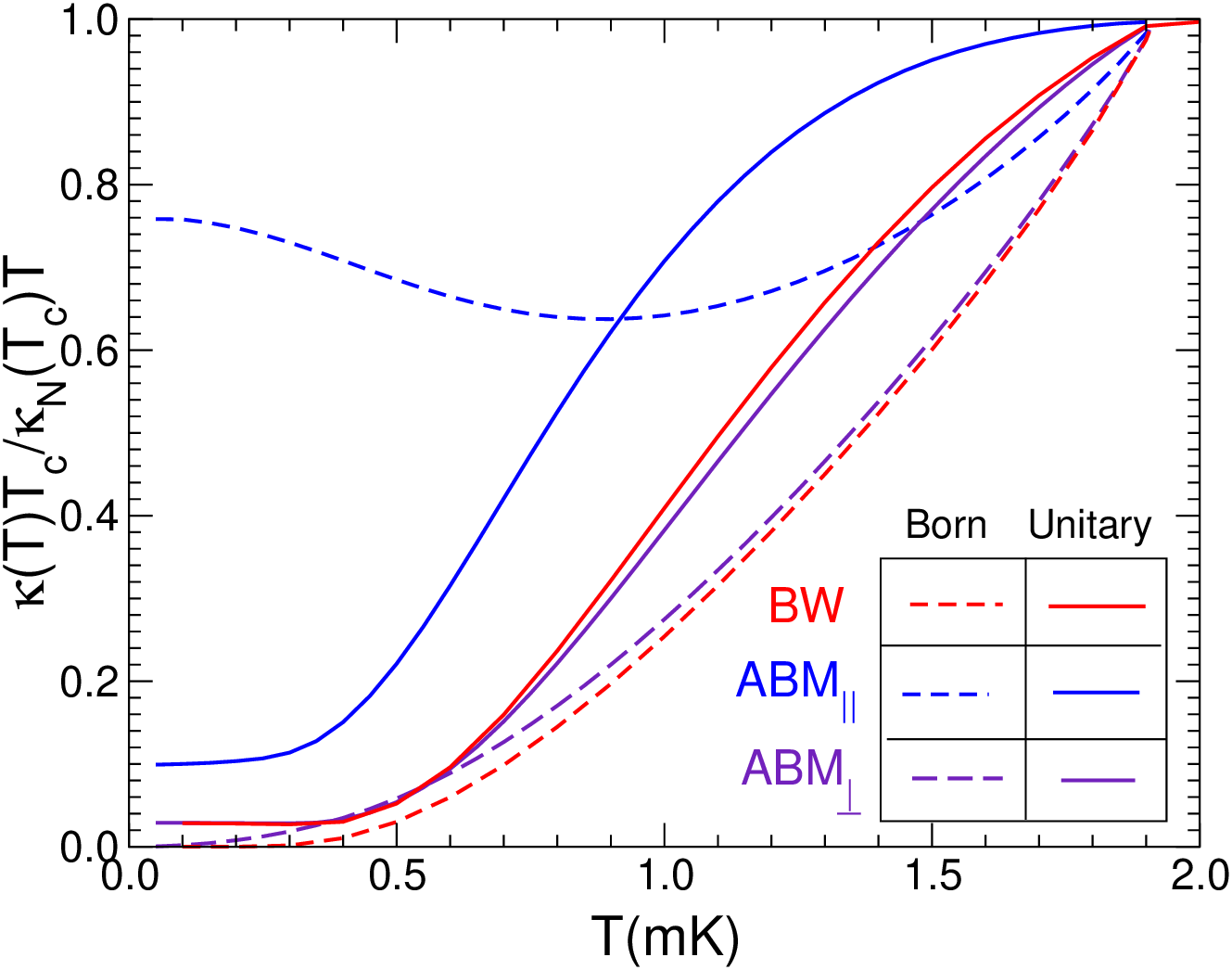}
\caption{Thermal conductivity for models of superfluid $^3$He in aerogel for Born and unitary
scattering at $p=30\,\mbox{bar}$.}
\label{Kappa_AB}
\end{center}
\end{figure}

The transition to the superfluid state is evident in Fig. \ref{Kappa_vs_LogT} as a change in the slope of the
thermal conductivity. Calculations of $\kappa/T$ for the ABM and BW-phases with aerogel scattering included
are shown in more detail in Fig. \ref{Kappa_AB} also for $\ell=180\,\mbox{nm}$. These results were obtained from
solutions to the quasiclassical
transport equations, following Graf et al. \cite{gra96a}, for spin-triplet, p-wave pairing.
Note the difference in 
the limiting $T\rightarrow 0$ behavior in the Born and unitary limits, and the sensitivity of $\kappa/T$ to the 
order parameter at $p=30\,\mbox{bar}$. However, in zero field the anisotropy of the thermal conductivity for the 
ABM state is likely to be averaged out by the orientational disorder of the $\vell$-texture.

At lower pressures, e.g. $p=10\,\mbox{bar}$ ($\xi_0/\ell = 0.16$), the excitation spectrum, and therefore the 
thermal conductivity, are expected to be less sensitive to the pairing symmetry. In the unitary limit
the spectrum of gapless excitations leads to a linear $T$-dependence of the thermal conductivity at low temperatures,
$k_BT\ll\gamma$, with a slope, $\lim_{T\rightarrow 0}\kappa/T$ that is determined by $\xi_0/\ell$.

\begin{figure}[ht!]
\begin{center}\leavevmode
\includegraphics[width=0.95\linewidth]{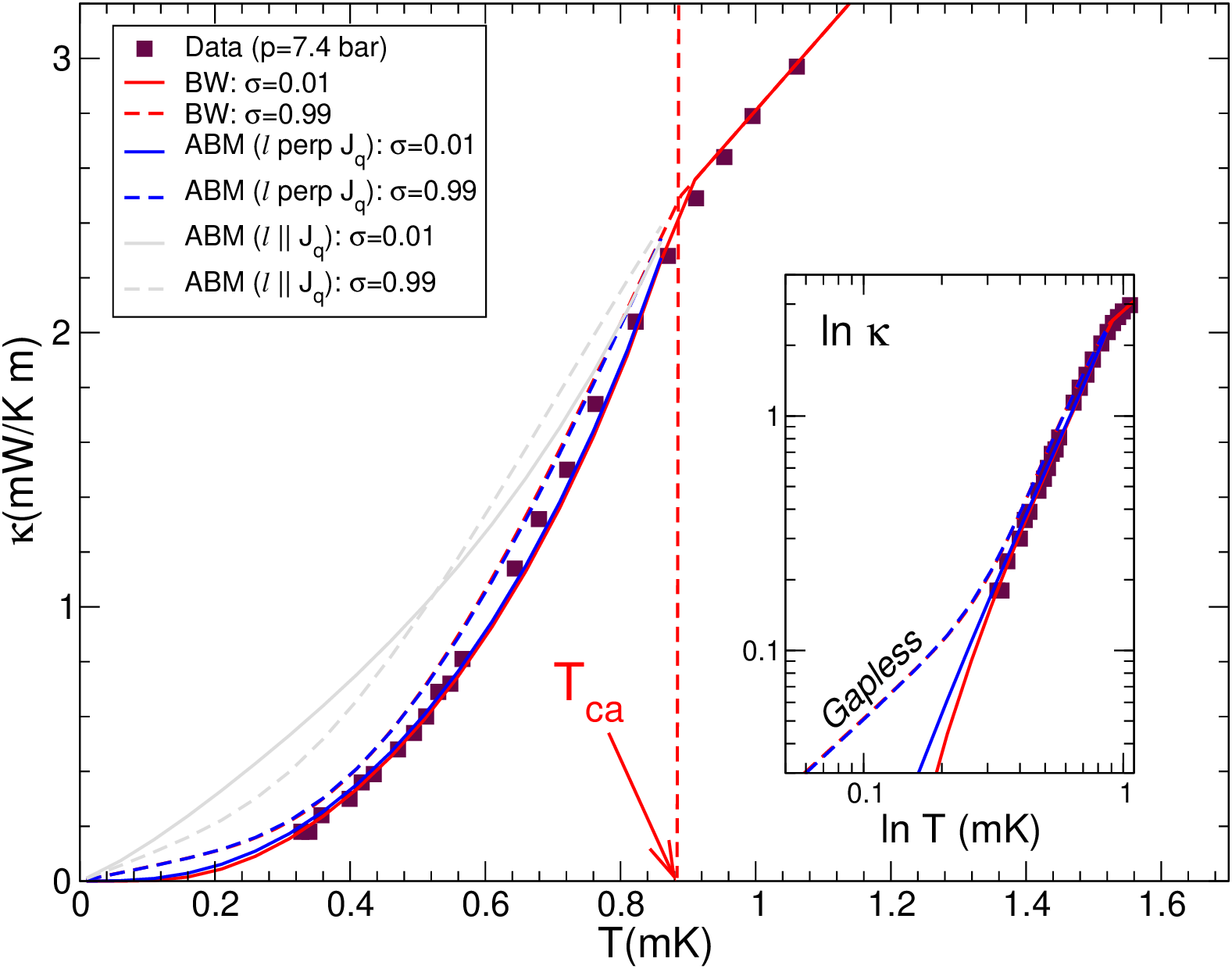}
\caption{The theoretical results for $\kappa$ (solid and dashed curves) are compared to data taken at 
Lancaster \cite{fis02} (squares) at $p=7.4\,\mbox{bar}$. The calculations are for the BW and ABM states
with a mean free path of $205\,\mbox{nm}$ in the Born ($\sigma=0.01$) and unitary ($\sigma=0.99$) 
scattering limits. Inset: Log-scale comparison between the unitary and Born calculations for the BW and ABM
state with $\vell\perp\vJ_{q}$.}
\label{Lancaster_Kappa}
\end{center}
\end{figure}

In Fig. \ref{Lancaster_Kappa} we compare our calculations to experimental data at $p=7.4\,\mbox{bar}$ from the 
Lancaster \cite{fis02} group. At this pressure pairbreaking is sufficiently strong that the thermal conductivity 
is only weakly dependent on the symmetry of the order parameter. The data, including $T_{ca}\simeq
0.88\,\mbox{mK}$ for $98\%$ aerogel, are accounted for by a mean free path of $205\,\mbox{nm}$ for either 
the BW phase or the ABM state with $\vell\perp\vJ_q$ (consistent with $\vB || \vJ_q$), in the Born limit. 
At this pressure the difference between unitary and Born scattering is significant only at very low 
temperatures, $T \le 0.2\,\mbox{mK}$ (inset of Fig. \ref{Lancaster_Kappa}). Thus, measurements at lower 
temperatures could provide evidence for gapless excitations ($\kappa\propto T$); measurements at higher 
pressures are more sensitive to the pairing symmetry.


\begin{thebibliography}{1}
\bibitem{fis02}
S.~N. Fisher, et al.,
\newblock {\em J. Low Temp. Phys.}, 126:673, 2001.
\bibitem{gra96a}
M.~J. Graf, et al.,
\newblock {\em Phys. Rev. B}, 53:15147, 1996.
\bibitem{sha00}
P.~Sharma and J.A. Sauls,
\newblock {\em Physica}, B 284-288:297, 2000.
\bibitem{thu98}
E.~V. Thuneberg, et al.,
\newblock {\em Phys. Rev. Lett.}, 80:2861, 1998.
\end{thebibliography}
\end{document}